\RequirePackage{fix-cm}
\documentclass[smallextended]{svjour3} 
\smartqed
\usepackage{graphicx}
\usepackage{multirow}
\usepackage{hhline}
\usepackage{comment}
\usepackage[dvipsnames]{xcolor}
\usepackage[colorlinks,bookmarksopen,bookmarksnumbered,citecolor=red,urlcolor=red]{hyperref}
\usepackage{hyphenat}
\usepackage{subcaption}
\usepackage{float}
\usepackage{caption}
\usepackage{fancybox}
\usepackage{array,booktabs}
\usepackage[toc,page]{appendix}
\usepackage{makecell}
\usepackage{rotating}
\usepackage{bigstrut}
\usepackage{booktabs}
\usepackage{subcaption}
\usepackage[misc]{ifsym}

\newcolumntype{P}[1]{>{\centering\arraybackslash}p{#1}}

\newcommand\rqone[0]{RQ1: To what extent do ML research repositories form the basis of other contributors' work?}
\newcommand\rqtwo[0]{RQ2: What are the types of changes in ML research repositories?}
\newcommand\rqthree[0]{RQ3: How do downstream changes differ from upstream changes in ML research repositories?}

\usepackage[graphicx]{realboxes}

\usepackage[most]{tcolorbox}
\definecolor{custom-gray}{cmyk}{0, 0, 0, 0.7, 1.00}
\newtcbtheorem[no counter]{Summary}{\hskip-0.97em}{enhanced,drop shadow={black!50!white},
 coltitle=white,
 top=0.15in,
 attach boxed title to top left=
 {xshift=1.5em,yshift=-\tcboxedtitleheight/2},
 boxed title style={size=small,colback=custom-gray}
}{summary}

\begin{document}
   \title {Towards a Change Taxonomy for Machine Learning Pipelines}
\subtitle {Empirical Study of ML Pipelines and Forks Related to Academic Publications}
 
\author{Aaditya~Bhatia \and
Ellis~E.~Eghan \and Manel~Grichi \and
William~G.~Cavanagh \and
Zhen~Ming~(Jack)~Jiang \and
Bram~Adams}

\institute{Aaditya~Bhatia (Ph.D. at SAIL) \and Bram~Adams (Director of MCIS Lab) \at
{ Queen's University\\
Kingston, ON, Canada}\\
\email{\{aaditya.bhatia, bram.adams\}@queensu.ca}\\
\and Ellis~E.~Eghan \at
{Assistant Professor, University of Cape Coast, Ghana}\\
\email{elliseghan@gmail.com}\\
\and Manel~Grichi \at
{Lead Data Scientist, VibroSystM Inc., Montreal, Canada}\\
\email{grichimanel@gmail.com}\\
\and William~G.~Cavanagh \at
{Student, Polytechnique Montreal, Canada}\\
\email{william.glazer-cavanagh@polymtl.ca}\\
\and Zhen Ming (Jack) Jiang \at
{Associate Professor and director of SCALE Lab, York University, Canada}\\
\email{zmjiang@cse.yorku.ca}\\
\and 
Aaditya Bhatia is the corresponding author\\
}

\date{Received: date / Accepted: date}
\authorrunning{Bhatia et al.}
\titlerunning{ML Code Changes}
\maketitle

\begin{abstract}
Machine Learning (ML) academic publications commonly provide open-source implementations on GitHub, allowing their audience to replicate, validate, or even extend the ML algorithms, data sets and metadata. 
However, thus far little is known about the degree of collaboration activity happening on such ML research repositories, in particular regarding (1) the degree to which such repositories receive contributions from forks, (2) the nature of such contributions (i.e., the types of changes), and (3) the nature of changes that are not contributed back to forks, which might represent missed opportunities. 
In this paper, we empirically study contributions to 1,346 ML research repositories and their 67,369 forks, both quantitatively and qualitatively, by building on Hindle et al.'s seminal taxonomy of code changes. 
We found that while ML research repositories are heavily forked, only 9\% of the forks made modifications to the forked repository. 42\% of the latter sent changes to the parent repositories, half of which (52\%) were accepted by the parent repositories.
Our qualitative analysis on 539 contributed and 378 local (fork-only) changes extends Hindle et al.'s taxonomy with two new top-level change categories related to ML (\textit{Data} and \textit{Dependency Management}), and 16 new sub-categories, including nine ML-specific ones (\textit{input data, parameter tuning, pre-processing, training infrastructure, model structure, pipeline performance, sharing, validation infrastructure, and output data}). While the changes that are not contributed back by the forks mostly concern domain-specific features
and local experimentation (e.g., \textit{parameter tuning}), the origin repositories do miss out on a non-trivial 15.4\% of \textit{Documentation} changes, 13.6\% of \textit{Feature} changes and 11.4\% of \textit{Bug fix} changes. 
\end{abstract}



\section{Introduction}
\label{sec:intro}

The notion of ``Software Engineering for Machine Learning/Artificial Intelligence'' (SE4ML/SE4AI) is becoming widespread in the software engineering community, with software engineering conferences featuring dedicated tracks and with dedicated venues appearing (RAISE, SEMLA, CAIN). The term ``machine learning software'' can mean different things depending on the context, spanning across a wide range of software projects from project-specific applications to third-party ML frameworks, to ML pipelines using such frameworks.

At one end of the spectrum, \textbf{Machine Learning (ML) frameworks} like Tensorflow\footnote{\url{https://www.tensorflow.org}} and PyTorch\footnote{\url{https://pytorch.org}} provide generic implementations of ML classification, regression, recommendation, and clustering algorithms, for use in any possible domain. At another end of the spectrum, \textbf{end user applications} integrate models into their code base to make domain-specific predictions. Since those models are domain-specific, an infrastructure is needed to continuously ingest data, perform pre-processing, build, tune and evaluate ML models specific to a given domain (e.g., image recognition or fraud detection), by orchestrating scripts and ML framework tools that produce datasets, models and evaluation/execution metadata~\cite{idowu21}. This infrastructure is called an \textbf{ML pipeline}~\cite{8804457,nahar22}, and forms the core of any organization's ML activities, catering to interdisciplinary teams of data scientists, data engineers, and developers~\cite{cd4ml}. An important subset of ML pipelines is produced in the context of published academic papers~\cite{fan2021makes}. Typically, researchers would upload a preprint of their work on ArXiv, including a GitHub repository with the pipeline code and (potentially) a labeled data set to train the pipeline on. Alternatively, other researchers or open-source developers might open-source an implementation or dataset of such a paper. Such code and data allow the open-source community to leverage state-of-the-art ML research to develop applications and benefit from the publications' ideas. Popular online indexes like PapersWithCode\footnote{\url{https://paperswithcode.com}} or ModelDepot\footnote{\url{https://modeldepot.io/}} provide searchable lists of papers, and their associated artifacts like implementations and/or datasets. Given the popularity\footnote{In August 2022, PapersWithCode indexed more than 77,640 academic AI publications along with their code bases.} of these ML pipeline open-source projects, the remainder of this paper focuses on this subset of ML pipeline projects.

While ML pipelines, including those related to an academic publication, typically are shared in the form of GitHub repositories\footnote{In the remainder of this paper, we use the term \textit{``ML research repositories''} to identify such  ML pipelines.}, an important question is to what extent such projects benefit from the open-source collaboration models leveraged by traditional (non-AI) GitHub projects, as opposed to just being an ``online backup'' or a ``replication package''. The typical GitHub collaborative coding model would see the OSS community \textit{fork} an ML research project \cite{zhou2020has}, make changes to the source code, and push those changes back to the original project using \textit{Pull Requests} (PRs). The developers of the original project would then check such PRs and accept or reject the proposed changes. Accepted PRs would then lead to code changes being merged in the ML research repository, a so-called \textit{upstream change}. Oftentimes, community developers could also make changes for their own use that they would not contribute back, i.e., so-called \textit{downstream changes}. GitHub's collaborative forking model has been known to improve the productivity of multifaceted software development and management tasks like making new features, handling issues, sharing knowledge, adding documentation, and managing upstream and downstream code~\cite{zhou2020has}.

While OSS collaborations in non-ML software are extensively studied by researchers~\cite{zhou2019fork,biazzini2014may,hu2016influence,lima2014coding,zhou2020has}, this is not the case in the context of ML. Certain assumptions of established software engineering activities like requirements engineering, software design, and quality assurance are no longer valid~\cite{washizaki2019studying,ozkaya2020really}, while a typical ML development team no longer only features traditional developer and tester roles, but also data scientists and data engineers~\cite{8804457}. Such multi-disciplinary collaboration leads to a wide range of different artifacts other than source code that require proper versioning and traceability with the code~\cite{idowu21}. 

Among the emerging software engineering practices replacing existing models of (multi-disciplinary) collaboration in the context of SE4ML, the types of code changes made on typical ML pipelines need to be explored to capture the nature of such community-based changes and compare them to pre-ML types of changes. In particular, in 2008, Hindle et al.~\cite{largecommits} presented a seminal taxonomy of code changes for traditional (non-ML) software, which identified seven code change dimensions, along with 24 sub-categories of changes (Table~\ref{tab:hindle}). Despite being 13 years old at the time of writing this paper, the taxonomy is still authoritative today. However, the advent of the ML era within the collaborative development environment calls for the need to substantially revise this taxonomy.

Hence, this paper empirically studies changes in ML research repositories and their forks, using a mixed-methods approach. First, we quantitatively mine the community collaborations to 1,346 ML research repositories (containing implementations of 1,144 arXiv publications) obtained via ModelDepot's ``Deep Search'' engine to analyze the behavior involving their forks, i.e., how active is online collaboration on ML research repositories? We then perform a large-scale qualitative analysis of 539 upstream and 378 downstream changes, adapting Hindle et al.~\cite{largecommits}'s taxonomy of code changes. Notably, we address the following research questions:
\begin{itemize}
    \item \textbf{\rqone} \\ 
      ML research repositories are heavily and transitively forked, yet overall only 9\% of forks made modifications. 41.6\% of the latter forks sent changes back to the parent ML research repositories (i.e., upstream changes), half of which (52\%) were accepted by the parent repositories. The time taken to merge those pull requests is about four times faster than for NPM packages on GitHub~\cite{dey2020pull} and 24 times faster than for GitHub projects in general~\cite{gousios2014exploratory}.
    
    \item \textbf{\rqtwo} \\ 
     Using the seminal code change taxonomy of Hindle et al.~\cite{largecommits} as a starting point, we identified two new categories of changes in ML research repositories, namely \textit{Data}, and \textit{Dependency Management}. Furthermore, we refined the taxonomy with 16 new sub-categories of changes. Nine of the sub-categories (i.e., \textit{input data, parameter tuning, pre-processing, training infrastructure, model structure, pipeline performance, sharing, validation infrastructure, and output data}) are ML-specific, while seven (i.e., \textit{add dependency, remove dependency, update dependency, file permissions, internal documentation, add auto-generated code, and program-metadata}) are more general sub-categories.
     
    \item \textbf{\rqthree} \\
    Manual comparison of changes contributed back by the forks to the origin ML repository (upstream changes) with changes not contributed back (downstream changes) shows that downstream changes typically are domain-oriented and add \textit{input/output data}, perform \textit{parameter tuning}, add new functional \textit{features}, and perform other non-functional changes like \textit{indentation, refactoring} or \textit{cleaning} up the source code. In contrast to this, upstream changes benefit the parent repository by \textit{updating dependencies} or \textit{fixing bugs} for the parent repository. Both downstream and upstream contributions add \textit{documentation}, and \textit{fix bugs}.
\end{itemize}
The remainder of the paper is organized as follows: Section~\ref{sec.data} presents the data collection and design of our study. Section~\ref{sec.related} summarises the related literature. Section~\ref{sec.results} discusses the motivation, approach, and results for each of our research questions. Section~\ref{sec.threats} discusses threats to the validity of our study while Section~\ref{sec.impl} explains the implications of our findings.  Finally, Section~\ref{sec.conclusion} concludes the paper.
\section{Related Work} \label{sec.related}

\subsection{Code Change Classification}
Prior work in code change taxonomies initiated in 1976 with Swanson et al.'s work on identifying changes during software maintenance in terms of \textit{corrective, adaptive} and \textit{perfective}~\cite{swanson1976dimensions}. The goal of such taxonomies originated from the need to enhance software decision-making. 

These changes were adopted as \textit{extended-Swanson categories} by Hindle et al.~\cite{largecommits} in the latter seminal taxonomy of software changes in 2008. A detailed description of Hindle et al.'s taxonomy is provided in Table~\ref{tab:hindle}. Despite its important role, the taxonomy is in need of updates. For instance, software development has become more collaborative since 2008 due to platforms like GitHub, leading to additional change types that would need to be added to the taxonomy in Table~\ref{tab:hindle}. Furthermore, the types of changes required in an ML setting like ML pipelines could lead to further missing change types, which this paper aims to study. Hence, our qualitative study builds on Hindle et al.’s change taxonomy, extending it with two new high-level change categories and 15 new sub-categories of changes.

    Later work shifted direction from establishing taxonomies to automated classification of code changes in terms of activities defined by such change taxonomies. In Hindle et al.'s~\cite{hindle2009automatic} later publication, the authors automatically classify maintenance changes into \textit{corrective, adaptive, perfective, feature addition}, and \textit{non-functional improvement} categories using ML techniques. Yan et al.~\cite{yan2016automatically} improved this approach of classifying code changes using a Discriminative Probability Latent Semantic Analysis (DPLSA) approach, which showed its benefits in multi-category classifications of code change activities during the evolution of software. Recently, in 2021, Ghadab et al.~\cite{ghadhab2021augmenting} further improvised the classification of code changes using BERT (Bidirectional Encoder Representations from Transformers) approach.
    
    Code change taxonomies are used for a variety of purposes. In 2009, Benestad et al.~\cite{benestad2009understanding} performed a literature survey on publications that assessed the impact of individual code changes on the maintenance and evolution of software systems. Wu et al.~\cite{wu2011relink} extracted missing links between bugs and committed changes by creating an automated tool, \textit{Relink}. Furthermore, Bissyandé et al.\cite{6498458} evaluated the efficacy of linking bug reports to code changes by benchmarking Relink against alternative bug-linking solutions. 
    Cortés-Coy et al.~\cite{6975661} used code changes to automatically generate commit messages. Farago et al.~\cite{farago2014impact} studied code changes to understand the impact of change operations (like add/update/delete) on ISO/IEC-9126 quality attributes of software. Software developers and researchers developing such tools may wish to inculcate our extended taxonomy of code changes to better support the development of ML systems.

\begin{table}[t]
\caption{Hindle et al.'s taxonomy of change types for traditional SE~\cite{largecommits}.
}
\label{tab:hindle}
\hskip-0.5cm\begin{tabular}{c|p{2.5cm}|p{8cm}}
\textbf{Category} & \textbf{Sub-Category} & \textbf{Definition} \\ \hline
\multirow{5}{*}{Maintenance}                                                                    & Bug fix               & Fixing bugs (e.g., adding exception control, conditional statements)                                                          \\ \cline{2-3} 
                                                                                                & Cross                 & Cross-cutting changes (e.g., logging)                                                                                         \\ \cline{2-3} 
                                                                                                & Maintenance           & Performing activities during maintenance cycle other than fixing bugs                                                         \\ \cline{2-3} 
                                                                                                & Parameter list change & Updating in the parameters list                                                                                               \\ \cline{2-3} 
                                                                                                & Debug                 & Setting up debug, tracking process (e.g., printing variable values, execution times)                                          \\ \hline
\multirow{4}{*}{Meta-Program}                                                                   & Documentation         & Changing the software documentation (e.g., read-me file, code comments)                                                       \\ \cline{2-3} 
                                                                                                & Build/Config          & Changing build or work-space configuration files (e.g., setup.txt or .yml)                                                    \\ \cline{2-3} 
                                                                                                & Testing               & Adding unit tests, bench-marking, changing test environment                                                                   \\ \cline{2-3} 
                                                                                                & Internationalization  & Adding language support other than English                                                                                    \\ \hline
\multirow{4}{*}{\begin{tabular}[c]{@{}c@{}}Non-Functional\\ Source Code \\ Change\end{tabular}} & Refactor              & Structural changes without changing the behavior (e.g., renaming variables, optimizing code)                                  \\ \cline{2-3} 
                                                                                                & Clean up              & Deleting code not used by the program (e.g., print statements, comments, unused imports)                                      \\ \cline{2-3} 
                                                                                                & Indent                & Adding proper indentation or formatting the code                                                                              \\ \cline{2-3} 
                                                                                                & Token replace         & Renaming tokens like variable or method names                                                                                 \\ \hline
\multirow{5}{*}{\begin{tabular}[c]{@{}c@{}}Source \\ Management\end{tabular}}                   & Merge                 & Merging commits or pull requests                                                                                              \\ \cline{2-3} 
                                                                                                & Source control        & Managing repository files (e.g, adding files to git ignore)                                                                   \\ \cline{2-3} 
                                                                                                & Versioning            & Changing the software release version                                                                                         \\ \cline{2-3} 
                                                                                                & Branching             & Creating a side development branch from the main branch                                                                       \\ \cline{2-3} 
                                                                                                & External              & Code submitted by developers who are not a part of the core team                                                              \\ \hline
\multirow{2}{*}{Implementation}                                                                 & Feature               & Adding new functional features                                                                                                \\ \cline{2-3} 
                                                                                                & Platform              & Changing hardware or platform-specific code (e.g., changing GPU hardware acceleration, changing file access for a new platform) \\ \hline
\multirow{3}{*}{\begin{tabular}[c]{@{}c@{}}Module \\ Management\end{tabular}}                   & Add module            & Adding modules/directories/files                                                                                              \\ \cline{2-3} 
                                                                                                & Move module           & Moving modules/directories/files                                                                                              \\ \cline{2-3} 
                                                                                                & Remove module         & Removing modules/directories/files                                                                                            \\ \hline
Legal                                                                                           & Licence               & Changing copyright or authorship                                                                                              \\ 
\end{tabular}
\end{table}

\subsection{Multi-repository software development via forking}
    Many researchers study collaborative development. For instance, Zhou et al.\cite{zhou2019fork} identified efficient practices for developers collaborating using forks. The authors build regression models to correlate efficient practices with respect to the behavior around forking. They found how the modularity of a code base and its contributions, as well as upfront management of which bugs require fixing by contributors, correlate with higher contribution volume and pull request acceptance. 

    Later, in 2020, in a follow-up work~\cite{zhou2020has}, Zhou et al. elucidated the perceptions around ``hard forks'' (forks that split development into a competing line of a new repository), against those of ``social forks'' (forks that create a public copy of the repository on a social website like BitBucket or GitHub). While hard forking traditionally has been considered a bad practice for developers and users~\cite{fogel2005producing}, the authors found that the perceptions around hard forking have changed in modern times.
    Nowadays, hard forks emerge out of social forks, and are seen as a positive non-competitive alternative to the original repository. 
    Constantino et al.~\cite{constantino2020understanding} identified the rationales, processes, and challenges behind collaborative activities on GitHub by conducting surveys. The authors found that GitHub collaborations contribute to software development, issue management, repository management, and documentation tasks.

    Brisson et al.~\cite{brisson2020we} studied collaborations on GitHub projects by analyzing transitive forks, user statistics, pull requests, and issues. Furthermore, Biazzini et al.~\cite{biazzini2014may} identified dispersion metrics for fork-induced code changes. Ren et al.~\cite{8449489} developed a web UI for the management of forking-based collaborations with features like fork searching and tagging. Other research~\cite{rahman2014insight,zhang2018multiple} studies the nature of upstream contributions in the form of \textit{Pull Requests} and identifies the nature of competing contributions. 
    
    However, none of this prior research studies the collaborative development of ML software. We build on the results from prior studies to compare the forking dynamics of ML research repositories.

\subsection{Software engineering for machine learning}
    ML systems are substantially different from traditional software systems and hence need dedicated research. For example, Washizaki et al. found that ML software engineering design patterns differ from those of non-ML software~\cite{washizaki2019studying}. Furthermore, in 2020, Ozkaya~\cite{ozkaya2020really} illustrated how the stochastic nature of ML changes the software development practices in ML. Overall, Martínez et al.~\cite{martinez2021software} performed a literature review on Software Engineering for AI-based systems.
    
    Recent widespread advances in ML have instigated researchers to study the maintenance activities and challenges in ML code. In 2019, Amershi et al.~\cite{8804457} uncovered the challenges in managing ML software at Microsoft, in particular identifying the typical ML pipeline and corresponding software activities. Furthermore, Zhang et al.~\cite{zhang2019empirical} and Arpteg et al.~\cite{arpteg2018software} studied the software challenges faced in deep learning applications. Sambasivan et al.~\cite{sambasivan2021everyone} identified data engineering challenges for which multiple roles of data engineers (like data collectors, annotators, ML developers, and data licensing teams) require powerful data infrastructure in order to support machine learning processes. In the context of the data lifecycle used for ML, Polyzotis et al.~\cite{polyzotis2018data} illustrated the challenges faced at Google. O'Leary and Uchida~\cite{o2020common} also studied problems with creating ML pipelines from existing code at Google.

    The work of Fan et al.~\cite{fan2021makes} is the closest to our paper. While the authors study a similar dataset as ours (i.e., 1,149 academic ML(AI) repositories referencing ArXiv publications), the authors focus on characterizing popular versus unpopular academic repositories in terms of the number of stars on GitHub, and analyzing factors correlations between the number of paper citations and GitHub repository metrics. However, our paper is the first to study the extent of actual OSS collaborations happening on ML research repositories (instead of paper activity based on those repositories), and to manually identify the nature of code changes performed on such ML pipeline projects.
    

\section{Data Collection and Experiment Setup} \label{sec.data}
    \begin{figure}[th]
      \centering
      \includegraphics[width=1\columnwidth, keepaspectratio]{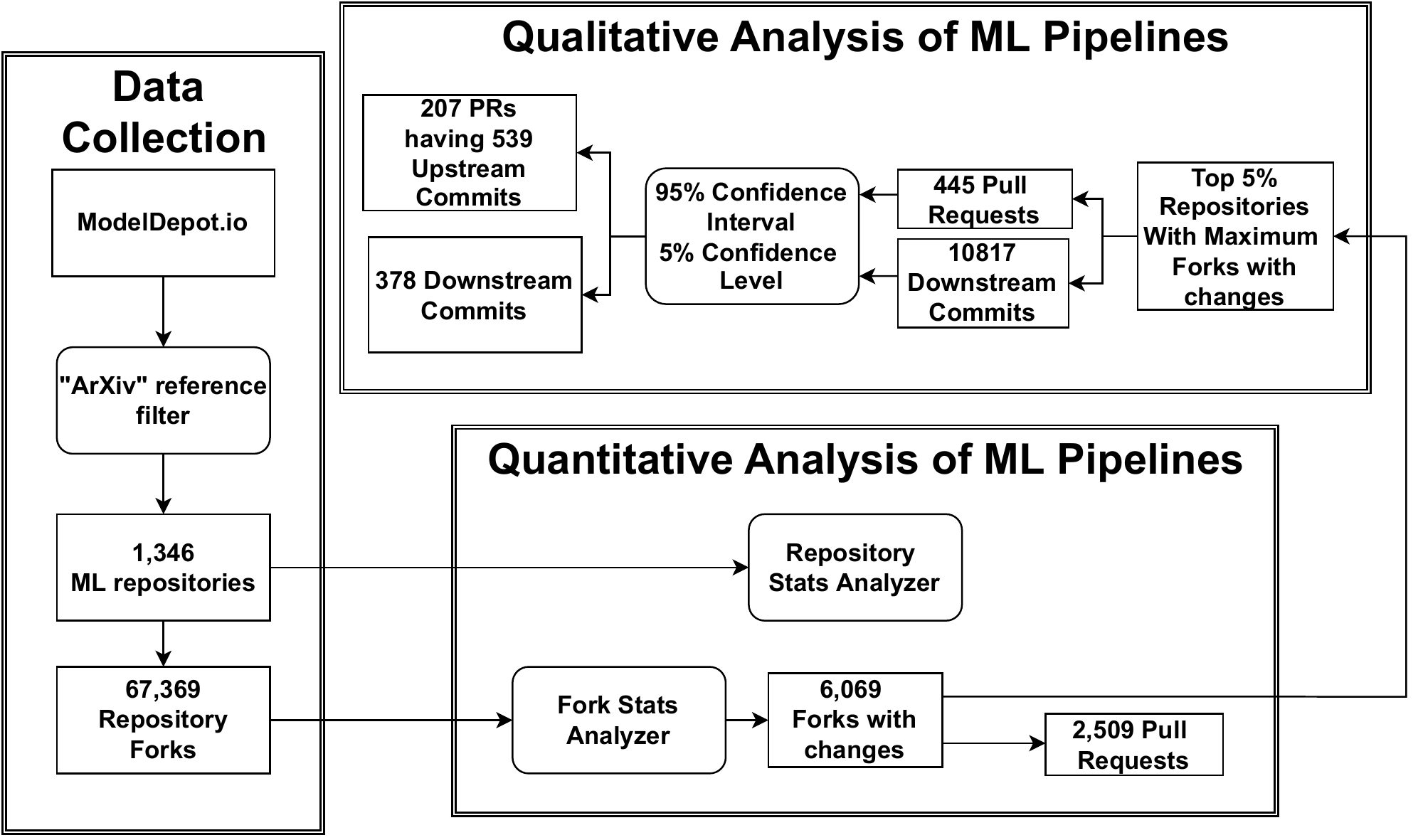}
      \caption{Data collection and processing steps.}
      \label{fig.datacollection}
    \end{figure}
    
    Figure~\ref{fig.datacollection} presents an overview of our data collection procedure along with the design of our empirical study to address the research questions of the introduction. RQ1 quantitatively studies the OSS collaboration characteristics on ML research repositories, while RQ2 and RQ3 perform a qualitative analysis on the nature of changes performed during this collaboration.
    
\subsection{Data Collection}
ModelDepot was a popular online model store containing 1) a catalogue of pre-trained ML models, and 2) a GitHub search engine for ML model pipeline implementations called ``Deep Search''. The latter engine effectively was an index of GitHub projects related to ML, allowing to search the projects based on name, ML framework (e.g., Tensorflow), programming language, and five model categories (i.e., ``computer vision'', ``natural language'', ``reinforcement learning'', ``generative'' and ``audio''). At the time this study was conducted in summer 2019, ModelDepot indexed over 50,000 ML implementations.

While both ModelDepot and its major competitor at that time, PapersWithCode, indexed GitHub repositories, we selected ModelDepot for our study because it was the most popular and diverse at the time of crawling the data (i.e., 50,000 model implementations compared to 8,500 on paperswithcode.com\footnote{\url{https://twitter.com/paperswithcode/status/1091315540092768257}}). The fact that ModelDepot did not require manual contributions to register new models, but leveraged its automated ``Deep Search'' engine to track new ML model repositories, was another reason why we opted for ModelDepot.
To collect the ModelDepot data, we built a scraper (crawler) to mine ML projects in the ``Computer Vision'' and ``Machine Learning'' categories, which were the two most common categories of models\footnote{\url{https://web.archive.org/web/20190404211946/https://modeldepot.io/search/results?q=}}. After sorting by the search engine's ``best match'' feature, we then focused on the top 5,000 non-fork projects.

Since this paper focuses on the evolution of ML model pipelines produced by researchers or inspired by the work of researchers, we filtered the 5,000 crawled projects using string matching to check the readme files for the presence of the term ``arxiv'' (referring to an ArXiv URL of an academic paper). This yielded 1,346 ML research repositories as our dataset for this study. Within this dataset, 1,144 unique academic papers were referenced and 23\% of these publications were referenced more than once. One of these publications, ``Deep Residual for Image Recognition''\footnote{\url{https://arxiv.org/abs/1512.03385}} was referenced the most (46) by the repositories in our dataset.

To further verify the soundness of our dataset, we did a separate analysis to check the quality of the studied ArXiv publications. To do so, from our dataset of 1,346 ML research repositories, we used a 95\% confidence interval and a 10\% confidence interval to obtain a sample of 94 repositories. For each of the 94 repositories, we checked whether their referenced ArXiv papers:
\begin{itemize}
    \item were peer-reviewed?
    \item involved authors from the industry?
    \item had any of the authors amongst the people contributing to the repository's development? 
\end{itemize}

To check whether publications were peer-reviewed, we looked at whether the ArXiv paper was also published at another venue, by leveraging \url{scholar.google.com}. Next, to check whether the authors were from the industry, we looked at the email addresses and the designations of the authors presented in the publication. Finally, we check whether the authors contributed to the repository.

As such, we found that all repositories from our inspected sample of 94 repositories are implementations of academic ML research. The ArXiv paper(s) referenced in the repositories' README file clearly showed the intention to implement the published algorithms, not just mentioning the work as a comparison or reference. Only \textbf{30\% of the repositories were developed by the referenced publications' authors themselves}, while the other other 70\% were implemented by other researchers or by open-source developers. We observed that the latter repositories basically referenced the paper(s) they were implementing. Since those repositories still represent an implementation of an ML publication shared with the open-source community, all of these represent genuine ML pipeline research repositories that we focus on in our research.

For instance, a repository named darkflow\footnote{\url{ https://github.com/thtrieu/darkflow}} referenced two academic publications\footnote{\url{ https://arxiv.org/pdf/1506.02640.pdf}}\footnote{\url{https://arxiv.org/pdf/1612.08242.pdf}} to indicate that they implemented the image processing algorithm YOLO, proposed by the said research. Another repository, ActivityRecognition\footnote{\url{ https://github.com/mohammed-elkomy/two-stream-action-recognition}} presented multiple publications\footnote{\url{ https://arxiv.org/pdf/1406.2199.pdf}}\footnote{\url{ https://arxiv.org/pdf/1604.07669.pdf}}\footnote{\url{ https://arxiv.org/pdf/1507.02159.pdf}} within the “Reference Papers” section of its ReadMe, for the same reason. While both repositories have no indication that one of the publications' authors was a contributor, the repositories clearly indicate that they implemented the referenced academic research.

Furthermore, \textbf{81\% of the repositories implemented peer-reviewed publications}, indicating that the implemented research is high quality. Finally, \textbf{59\% of the repositories referenced publications involving authors from the industry}, indicating how repository development focuses on industry-relevant research.

\subsection{Quantitative Analysis of Forking in ML Research Repositories (RQ1)} To analyze the dynamics of OSS collaboration through forks of ML research repositories (RQ1), we use the GitHub Search API\footnote{\url{https://docs.github.com/en/rest/reference/search}} to analyze each of the 1,346 repositories in our dataset. We perform our quantitative analysis via seven metrics related to forking as described in Table~\ref{tab:metrics}. Three out of these metrics (the bolded ones in Table~\ref{tab:metrics}) are adapted from Brisson et al.~\cite{brisson2020we}.

Given a large number of forks, Brisson et al.~\cite{brisson2020we} suggested that fork data is noisy. For this reason, we introduce the concept of \texttt{Forks\_with\_changes} to identify \textit{forks with modifications}. 
Such \texttt{Forks\_with\_changes} contain at least one commit that does not occur in the parent repository (downstream changes), or contributed back at least one commit via a pull request (upstream changes). 

To identify forks with downstream changes, we first calculate for each fork $F$ the set of commits $S_{F}$ whose commit id does not occur in the parent repository. Since the resulting set $S_{F}$ of fork commits could still contain commits that have been merged upstream through rebasing (changing their commit id), we then check for each commit in $S_{F}$ to see if any commit in the parent repository has the same commit message subject, author name, and author date, since those metadata fields have been found to be stable during rebase~\cite{german2016continuously}. If so, we remove those commits from $S_{F}$, since they also exist in the parent repository. Since we may be missing cases where a fork had all of its commits merged as PRs, we then check the list of forks that submitted PRs using the GitHub Search API, and add such forks into $S_{F}$. If the resulting $S_{F}$ is not empty, we consider each $F$ in  $S_{F}$ to be a \textit{changed fork}. 

We used the \texttt{Star\#} and \texttt{Fork\#} metrics to measure the popularity of repositories, as indicated by Borges et al.~\cite{borges2018s}. We computed the \texttt{First\_Fork\_time} and \texttt{Final\_fork\_time} metrics to indicate the temporal aspects of ML research repositories. The \texttt{First\_Fork\_time} indicates the speed of the OSS community in adapting an ML implementation, whereas the final fork time indicates the longevity of collaboration activities on ML pipeline code.

In addition, we calculate for each repository the number of forks, both direct and transitive, as a measure of the value of these repositories for the OSS community in terms of collaborative potential. We call the direct forks of a repository \emph{level-one transitivity}, while a transitivity of \emph{level-two} indicates the transitive forks of the direct (level-one) forks, and so on. The higher the proportion of repositories with at least one \emph{level-two} fork, the more collaboration seems to happen.

\begin{table}[t]
\caption{Metrics used in RQ1. The bolded metrics are adapted from Brisson et al.~\cite{brisson2020we}.}
\label{tab:metrics}
\begin{tabular}{|l|p{8.8cm}|}
\hline
Metric & Definition \\
  \hline
\textbf{Star\#} & \# users who starred the repository. \\
\textbf{Fork\#} & \# forks created from the repository.\\
Forks\_with\_changes\# & \# forks where the forked code base is modified. \\
\textbf{PR\#} & \# PRs sent to the parent repository. by its forks. \\
PR\_Accept\% & (\# PRs merged into parent repository) /  (PR\#) \\
First\_Fork\_Time & \# Days between creation of a repository, and its first fork. \\
Final\_Fork\_Time & \# Days between creation of a repository, and its final fork. \\
\hline
\end{tabular}
\end{table}

\subsection{Qualitative Analysis of Change Types in Forks (RQ2/RQ3)}
\label{sec:qualapp}

Since ML software has obtained a prominent place in software engineering, and the nature of the machine learning software lifecycle is substantially different from that of traditional software~\cite{8804457}, one would expect further change types to be added. ML practitioners use data pre-processing techniques, iteratively \textit{tune} model hyperparameters for obtaining the most optimal ML model under the data science life cycle~\cite{8804457,nahar22}. Reusing ML software requires users to understand the rationale behind the ML implementations, instigating users to change their code for \textit{documenting the internal working }of the ML code. All of this has led to a variety of types of artifacts other than source code that require changes as well~\cite{idowu21}.
In this study, we build on Hindle et al.'s taxonomy to identify the types of changes in ML research repositories by studying a statistically significant sample of 1) fork PRs merged by the parent ML research repository (i.e., \textbf{upstream changes}) and 2) commits within the forked repositories that were not submitted as PRs (i.e., \textbf{downstream changes}). We describe our qualitative analysis process below:

\begin{enumerate}

    \item  \textbf{Selection of repositories for sampling.}
    
      Since not all forked repositories made changes, let alone sent them as PRs upstream, we first select the repositories having the most active forks. To do so, we ordered the repositories by our metric \texttt{Forks\_with\_changes} and obtained the top 5 percentile, i.e., 23 repositories. Overall, these 23 repositories had 10,817 downstream commits and 445 PRs. From these, using a 95\% confidence level and 5\% confidence interval, we obtain a statistically representative sample of 1) 207 PRs containing 539 upstream commits (\textbf{upstream sample}\footnote{PRs on GitHub are not limited to just one commit.}), and 2) 378 downstream commits (\textbf{Downstream Sample}).

      Both samples were stratified, such that repositories with a higher proportion of \texttt{Forks\_with\_changes} had more data points in the samples.  
      Since a PR can comprise multiple commits, we selected all the commits for the sample of 207 PRs and obtained 539 upstream commits. Similar to obtaining the stratified upstream sample, we used the proportion of \texttt{\#Commits} with respect to the \texttt{Forks\_with\_changes} of each of the 23 parent repositories for creating the stratified downstream sample.

  \item \textbf{Study Participants.}
  We used teams from two universities to manually classify the types of changes in upstream and downstream commits. \texttt{University-A} classified downstream changes while \texttt{University-B} classified upstream changes. Both teams included two or more grad students, and one faculty member, all having knowledge of ML and non-ML software design and development. Due to the large-scale nature of our study, we employed four coders in \texttt{team A} and three coders in \texttt{Team B}.
  
  \item \textbf{Extending taxonomy of changes.}
    From the sample of 378 downstream commits, \texttt{Team-A} first performed a pilot study on an initial sample of 78 commits to validate the extent to which Hindle's taxonomy~\cite{largecommits} was able to classify code changes or required refinements. The 78 commits were distributed across the three coders of \texttt{Team-A} such that each commit had two coders and each coder had 26 commits in common with each of the other coders. The assignment of commits to each coder was anonymous.

    Each coder then individually coded their 52 (2 x 26) assigned commits, identifying all types of changes within the commits under study (more than one type of change could apply). In cases where the change (sub-)type could not be found using Hindle's change categories, the coders individually could create a new category. Once finished, the coders met online discussing only the new types of changes that they had identified, without considering the specific commits tagged with these new types. The proposed new types could be merged, renamed, or removed until a consensus was reached. 
    
    \texttt{Team-A} then 
    re-labeled their samples using the enhanced 
    Hindle's taxonomy. Once done, the coding results were combined into a spreadsheet. In the first phase, each coder had to check the commits they were assigned that had conflicting coding results. This was done asynchronously by adding comments on the spreadsheet. If a coder was in accord with the other coder's interpretation, a disagreement was resolved; otherwise, it was left open. In a second phase, the remaining disagreement cases were then discussed in person by \texttt{Team-A}, possibly refining the taxonomy.

    With this final version of the taxonomy, \texttt{Team-A} started labeling the remainder of its samples, using the same style of assignment as for the initial 78 (i.e., anonymously sharing the same number of commits with each other coder). In parallel, \texttt{Team-B} was assigned the 207 PRs in a similar manner. While both teams could still make changes to the taxonomy during this coding, we observed saturation in the labels after tagging the initial set of 78 downstream commits, i.e., no new (sub-)categories were identified in the later part of labeling the 300 downstream samples and 539 upstream samples.

  \item \textbf{Calculation of inter-rater agreement.}
    Once coding was finished, both teams individually used the spreadsheet-based and in-person resolution of disagreements used initially for the first 78 cases. To calculate inter-rater agreement, since each sample was rated by two participants, we use Krippendoff's Alpha~\cite{krippendorff2011computing} as our metric for the inter-rater agreement. This metric supports multi-label classification by multiple participants. A similar approach of obtaining inter-rater agreement in a multi-rater setting was used by Heng et al.~\cite{li2020qualitative} to manually tag logging data, and Salza et al.~\cite{salza2018developers} to classify mobile app updates.

    Across the three coding activities (78 and 300 commits for \texttt{Team-A}, and 207 PRs for \texttt{Team-B}), the teams reported high agreements with a Krippendoff's $\alpha$=98\% on the sample of 378 downstream changes and a Krippendoff's $\alpha$=92\% for the sample of 539 upstream changes. These values of inter-rater agreements are high and reflect the statistical robustness of our data labeling results. 

   \end{enumerate}

   Given that the final change taxonomy spans 39 change sub-categories, and that we coded 378 downstream commits and 207 upstream PRs (containing 539 commits) across two teams of seven coders (and two universities), the resulting empirical study was non-trivial. For instance, in the downstream change\footnote{\url{https://github.com/BoseAslCohort/youtube-8m/commit/c1b01315bafc24e83248cd862a9324bb21d4d52d}} performed by fork ``BoseAslCohort'' for the project ``Youtube-8m'', the authors manually inspected changes for 27 changed files, which included 11,755 code additions. Overall, it took an estimated six man-months to finish the qualitative study.

\section{Case Studies}\label{sec.results}
\subsection{\rqone}

\noindent{\bf Motivation.} Currently, there is no empirical evidence regarding the extent to which 1) open-sourcing ML research code helps the OSS community in building new applications and 2) the OSS community contributes and helps maintain the original ML research implementations. In contrast, for non-ML software, prior research~\cite{zhou2019fork,biazzini2014may,brisson2020we,rahman2014insight,zhang2018multiple} has studied the nature of multi-repository development and maintenance of OSS projects. Hence, in this RQ, we analyze the OSS development activities around research-based ML pipeline repositories.

\noindent{\bf Approach.}
As discussed in Section~\ref{sec.data}, we extract 1,346 GitHub repositories having references to machine learning ArXiv publications, then use the GitHub Search API\footnote{\url{https://docs.github.com/en/rest/reference/repos}} to obtain the metrics identified in Table~\ref{tab:metrics}. \\

\noindent{\bf Results.} 
\subsubsection{Forks\_with\_changes}
\begin{figure}[t]
    \includegraphics[width=\linewidth]{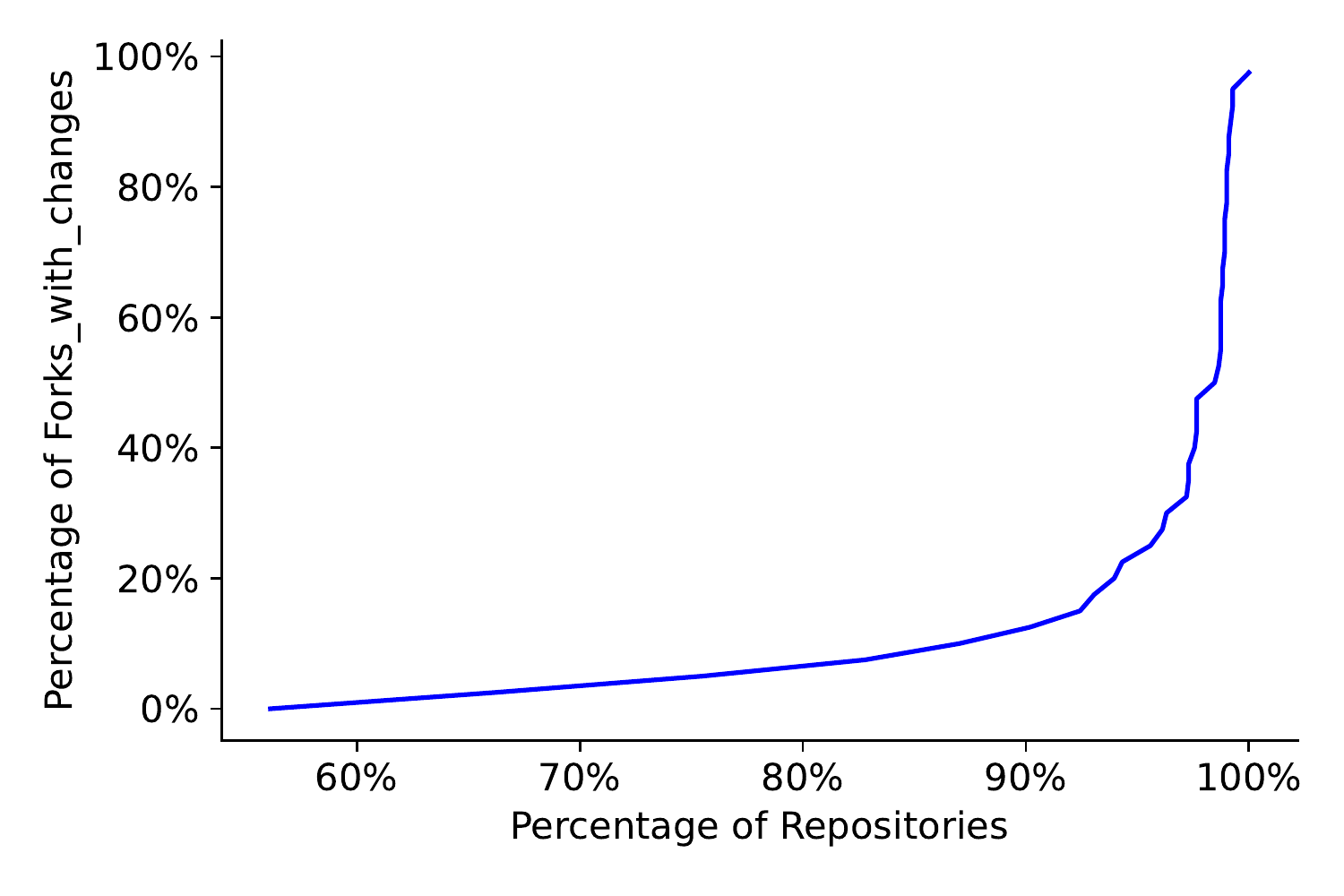}
    \caption{
    Percentage of \texttt{Forks\_with\_changes} across studied repositories. 52\% repositories do not have any modifications to the forked source code.}
    \label{fig:useful_fork_prcntage}
\end{figure}
\textbf{Only 9\% of forks of the ML research repositories have modifications to the forked source code (i.e., have \texttt{Forks\_with\_changes})}. Overall, 82.5\% of the 1,346 repositories had forks. Figure~\ref{fig:useful_fork_prcntage} shows the cumulative percentage of those repositories with a fork having at least one \texttt{Fork\_with\_changes}. Since 51.6\% of the repositories do not have any fork modification (only non-changed forks), and the slope of the curve is gentle and linear until 90\%, the percentage of ML research repositories with \texttt{Forks\_with\_changes} is low. \\

\subsubsection{Popularity of ML research repositories}
\begin{figure}[t]
    \centering
    \includegraphics[width=\columnwidth, keepaspectratio]{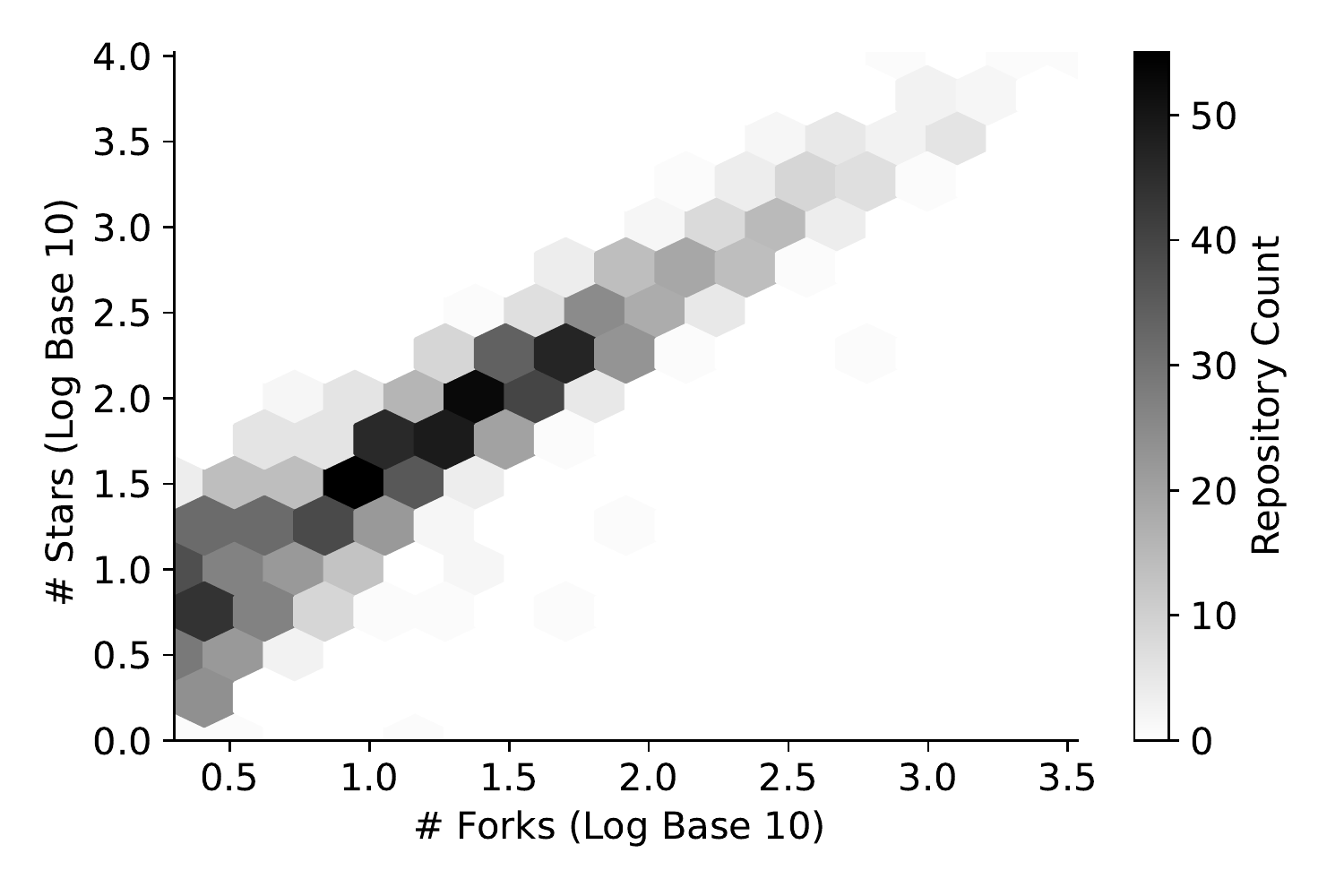}
    \caption{Hexbin for \texttt{Star\#} (logged) and \texttt{Fork\#} (logged) indicates high correlation. Popularity can be indicated by either of the metrics.}
    \label{fig:repo-stars-forks}
\end{figure}
\textbf{ML research repositories have a high median \texttt{star\#} of 22 and \texttt{fork\#} of 8.} 
These numbers stand in stark comparison to the datasets used by prior research for non-ML repositories. We observed a median of zero stars and forks for the replication dataset provided by Brisson et al.~\cite{brisson2020we}, consisting of 13,431 projects. The comparisons of \texttt{star\#} and \texttt{forks\#} for Brisson's dataset with our study are statistically significant with Wilcoxon Rank sum $p-value<0.01$ \\
\textbf{\texttt{Star\#} and \texttt{fork\#} are highly correlated (Spearman $\rho$=0.94)}
as shown by the data distribution in Figure~\ref{fig:repo-stars-forks}. As indicated by the darker color at the low end of \texttt{star\#} and \texttt{fork\#} in Figure~\ref{fig:repo-stars-forks}, 20\% of the repositories have less than five stars and five forks.

These results again contrast to the low correlation of 0.45 found by Brisson et al. on their non-ML dataset of 13,431 repositories, suggesting a much weaker connection between stars and forks. Several hypotheses might explain this contradiction, and require future work to be validated. For example, due to the current hype of AI technologies, ML repositories might be substantially more popular than non-ML repositories. It could also be that, due to the quick succession of new AI algorithms, the OSS community uses forks for the purpose of ``bookmarking'' or keeping copies of interesting ML research implementations~\cite{Kalliamvakou14}. One indication of the latter hypothesis could be the high percentage (91\%) of forks without any code change (i.e., non-changed forks) that we found earlier.
\begin{figure}[t]
    \centering
    \includegraphics[width=\columnwidth,keepaspectratio]{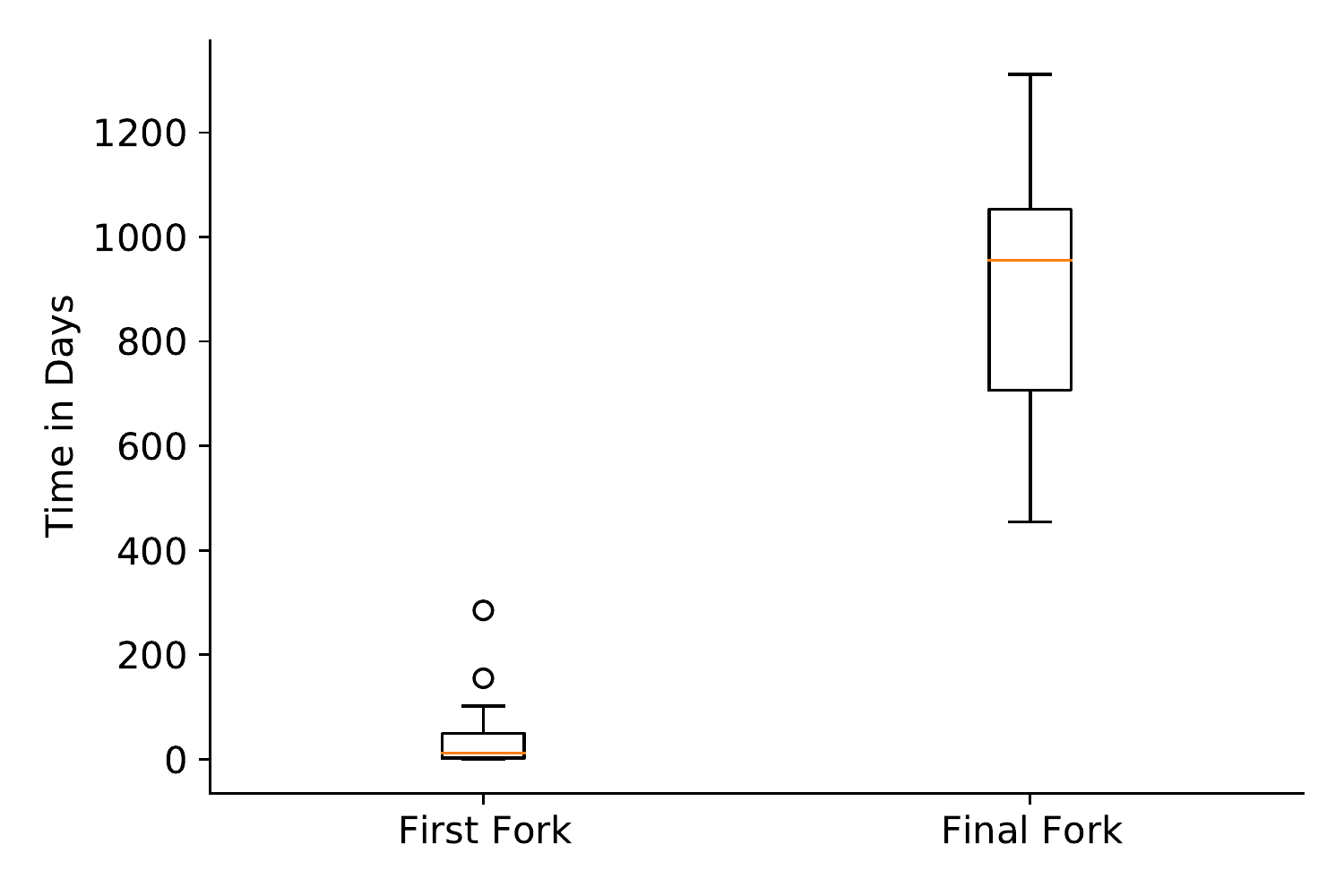}
    \caption{The first forking time represents the speed of the open source community in adapting ML research repositories, whereas the final forking time represents the longevity of making contributions to the ML repositories.}
    \label{fig:fork_minmaxtimes}
\end{figure}
\subsubsection{Speed and Longevity of Forking}
\textbf{Forks on ML repositories appear as fast as the 11\textsuperscript{th} day (median 11.5 days), while fork-based collaboration sustains a median of 2.6 years.} Figure~\ref{fig:fork_minmaxtimes} shows the distribution of the time of the first and the final (at the time of analysis) fork for each repository in our dataset. A median ML research repository receives its first forks on the \textbf{11\textsuperscript{th} day} after creation date,
while an ML research repository is forked till a median of \textbf{2.6 years} of the creation of the repository. Although the \texttt{final\_fork\_time} may be impacted by the time of our analysis, nevertheless, a value of 2.6 years definitely shows that the ML repositories are not just data dumps but can foster online collaboration.

\subsubsection{Transitive Forking}
\textbf{Transitive forks (i.e., fork repositories with their own forks) are present in 20\% of ML research repositories.}. We observed 67,369, 1,581, 44, and 7 cases of direct forks, level-two, level-three, and level-four forking transitivity for 1,110, 226, 28, and 3 ML research repositories respectively. Hence, 226 out of 1,346 repositories, i.e., 20\%, have at least one transitive fork (since the 28 with level-three forks are a subset of the 226, etc.). The ML research repository with the highest forking transitivity in our dataset includes the Autopilot-TensorFlow project\footnote{\url{https://github.com/SullyChen/Autopilot-TensorFlow}}, which is an implementation of self-driving car research\footnote{\url{https://arxiv.org/pdf/1604.07316.pdf}}. This project has 281 direct forks, 10 level-two, 3 level-three, and 5 level-four transitive forks.

We compare our findings to the fork transitivity results reported by Brisson et al.~\cite{brisson2020we}, who conducted an analysis of the March 2019 GHTorrent dataset and reported 
12,171 level-one, 778 level-two, 84 level-three, 11 level-four, and 2 level-five forks. A $\chi^2$ test of independence between these findings and ours yielded a $p-value<0.01$, representing a statistically significant difference in data distribution between ML and non-ML repositories.

\subsubsection{Upstream Contribution}
\textbf{In terms of upstream contributions to the ML research repository, 41.6\% of \texttt{Forks\_with\_changes} send changes back to the original repositories } A total of 607 pull requests were submitted upstream by \texttt{Forks\_with\_changes}, out of which 316 were merged into the original repositories. This resulted in \textbf{52.1\% acceptance.} This value is slightly lower than that of a recent study on NPM packages by Dey and Mockus~\cite{dey2020pull}, who reported a PR acceptance rate of 60\%.

\begin{figure}[t]
    \centering
    \includegraphics[width=\columnwidth, keepaspectratio]{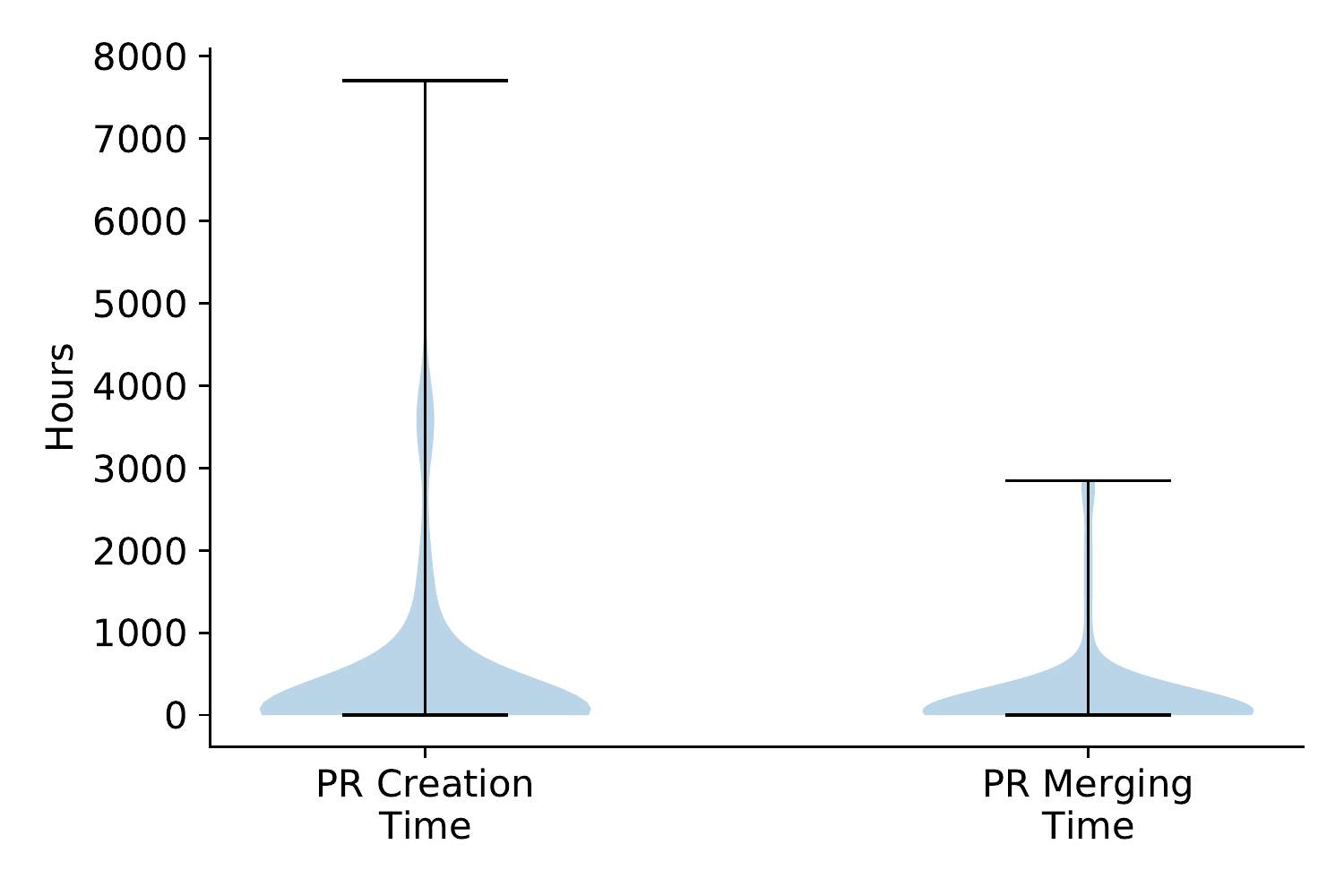}
    \caption{Time spent in sending PRs (left) and merging the PRs into the upstream parent repository (right).}
    \label{fig:violins}
\end{figure}

\textbf{27.5\% of the upstream PRs were submitted on the same day as that of the creation of the fork}. In particular, a fork takes a median of 22 hours to submit a PR. After receiving a PR, the parent ML research repository takes a median of seven hours to review the upstream changes before deciding on them, as shown by the violin plots in Figure~\ref{fig:violins}. This is approximately four times faster than the median PR acceptance times (27.7 hours) for NPM packages on GitHub~\cite{dey2020pull}. Another study performed on 1.9 million PRs on GitHub in 2013 by Gousios et al.~\cite{gousios2014exploratory} reported a median of seven days to merge a PR.

\begin{Summary}{Summary of RQ1}{}
The OSS community forks the ML research as fast as a median of 11.5 days after the creation of the repository, and forking continues until a median of 2.6 years. ML research repositories are heavily and transitively forked, yet only 9\% of the forks have modifications. Of those, 43\% sent upstream changes back to the parent ML research repositories, with a 52\% acceptance rate. PR merge times of ML research repositories are faster than the values reported by prior research for non-ML repositories.
\end{Summary}
\label{subsec.RQ1}

\subsection{\rqtwo}\label{subsec.rq2}

\begin{figure}[t]
    \centering
    \includegraphics[width=1.2\columnwidth,keepaspectratio]{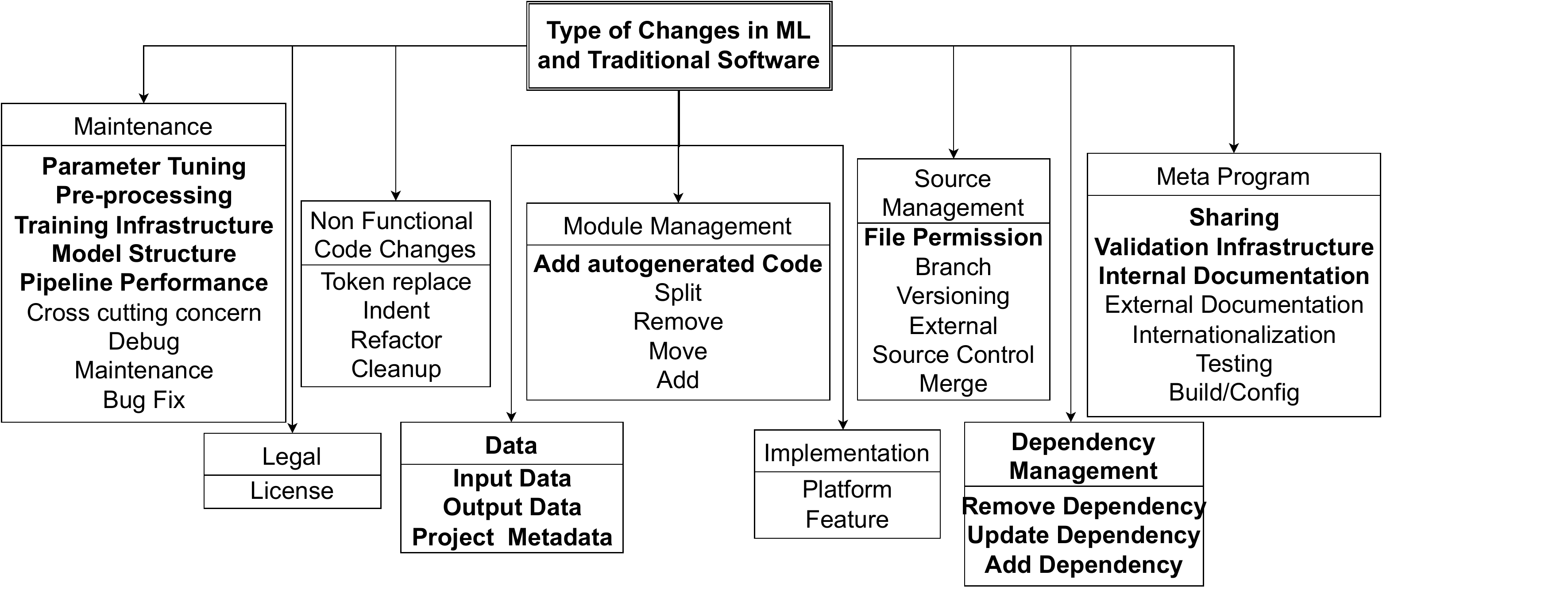}\caption{Enhanced version of Hindle's change taxonomy. The bolded change (sub-)categories were identified by this study. }
    \label{fig:Finaltaxonomy}
\end{figure}

\noindent\textbf{Motivation.} 
Since Hindle et al.'s taxonomy of changes focused on traditional software systems known in 2008~\cite{largecommits}, this research question performs a qualitative analysis to identify the types of changes made in ML research repositories, possibly extending Hindle's change taxonomy. Through this, we wish to help software practitioners in building and maintaining ML software, which not only involves code changes but also changes to many other kinds of artifacts~\cite{idowu21} (e.g., dataset and models). Furthermore, training/education teams need an understanding of the types of code changes to better equip students and novice developers in supporting ML applications.

\noindent\textbf{Approach.}
In \textbf{RQ1}, 
we observed that 52\% of the PRs sent by \texttt{Forks\_with\_changes} are merged into the parent ML research repository. In this research question, we qualitatively analyze 1) the types of changes that were merged with the parent repositories, which we call \textbf{upstream changes}; and 2) the types of changes that were performed within the forked repositories, but not pushed to the parent repository, which we call \textbf{downstream changes}. Using the sampling and coding approach discussed in Section~\ref{sec:qualapp}, the coders of \texttt{Team A} and \texttt{Team B} validated and enhanced Hindle et al.'s taxonomy~\cite{largecommits}. This section reports on the new change (sub-)categories identified in the analyzed code changes of ML repositories.\\

\textbf{Results.} 
\textbf{Hindle et al.'s change taxonomy~\cite{largecommits} was extended with two new categories and 16 new sub-categories of changes.} 
Only one of the two new high-level change categories was ML-specific, i.e., \textit{Data}, while the other one, i.e., \textit{Dependency Management}, represents an update to the original taxonomy related to modern library dependency management activities (which may have been less relevant 13 years ago). 
A graphical summary of the extended taxonomy of change (sub-)categories is provided in Figure~\ref{fig:Finaltaxonomy}. In the subsections below, we briefly describe and illustrate each new (sub-)category and how it complements the existing taxonomy:

\subsubsection{Maintenance}
Code changes performing software maintenance activities. We identified four new maintenance change sub-categories in the context of machine learning.

\begin{itemize}
 \item \textbf{Pre-processing:}\\
 \textit{Definition:} Source code changes related to manipulation, cleaning, and filtering of data before feeding it to the model training or inference components of an ML pipeline.\\
 \textit{Explanation:} 
 ML model training needs high-quality, clean data~\cite{ng2021mlops} making data pre-processing a vital part of the ML pipeline~\cite{8804457}. Pre-processing changes internally mutate and clean the ingested data such that it can be consumed by the ML model, for example by changing the text-embeddings for NLP data. In contrast to the \textit{Data category}, which deals with the ingestion or egestion of external data to/from the ML pipeline, pre-processing changes deal with internal data manipulation, and hence is a maintenance activity.\\
 \textit{Notable Instances:} In an image processing application~\cite{prep_eg}, pre-processing changes involve changing image color formats (Grayscale, RBG, BGR) before feeding the images to the ML model. \\
 
    \item {\textbf{Parameter Tuning}} \\
    \textit{Definition:} Changes made to hard-coded (hyper-)parameter values for tweaking the performance or functionality of an ML pipeline.\\
    \textit{Explanation:} A machine learning pipeline consists of many different phases (e.g., data preprocessing, feature extraction, model training, and validation) that, together, aim to generate models with the best fit possible. Each of these phases~\cite{8804457} involves choosing values for various thresholds, model hyper-parameters, and other configuration options, many of which have hard-coded values. Hence, ML pipeline developers often find themselves tweaking these hard-coded values 
    while building the model or preprocessing the data (i.e., parameter tuning). Even in the case of model hyper-parameters, which are optimized during the model training process, their initial (hard-coded) value or range often has to be chosen well for quicker convergence during training. In contrast to \textit{model structural} changes, which change the model building code at a structural level, parameter changes are performed at the variable (value) level.\\
    \textit{Notable Instances:} 
    The model hyper-parameter variable, \textit{weights\_regularizer} in project \texttt{youtube-8m}~\cite{param_tun}, was changed from $1e^{-5}$ to $1e^{-8}$. In another image processing application~\cite{param_tun_eg2}, adding a sliding window variable for pre-processing of video frames is a parameter tuning change.\\
    
    \item \textbf{Model Structure:}\\
        \textit{Definition:} Structural change to the source code responsible for training the machine learning model.\\
        \textit{Explanation:} Model structural changes involve changing the code encompassing the structure of deep learning models or the various functions or modules that manipulate the model during training, such as adding functions for dropout layers, loss functions, or regularizers for a model class. These changes are different from \textit{parameter tuning} changes since they are at a structural level rather than the variable level.\\
      \textit{Notable Instances} In file \texttt{train\_val.py} of \texttt{tf-faster-rcnn} project\footnote{\url{https://github.com/shikorab/tf-faster-rcnn}}, the model structure was changed to accommodate features related to image and mask height and width, which manifested in numerous structural changes to the model building process~\cite{mod_tra}.\\

    \item \textbf{Training Infrastructure:} \\
    \textit{Definition:} Pipeline-level changes performed for training the model.\\
    \textit{Explanation:} Amershi et al.~\cite{8804457} identified the canonical components of a typical ML pipeline, such as data wrangling, feature engineering, and model training. Making changes in one component (for instance, a different dataset schema) may manifest in other pipeline components (e.g., data pre-processing, feature engineering, model training, etc.). As such, training infrastructure changes correspond to any pipeline-level changes to the logic driving the ML model training phases. This is similar to how Hindle et al.~\cite{largecommits}'s original Build/Config change sub-category  focuses on the logic driving the compilation (build) process, in contrast to changes to the actual source code (in our case: changes to, for example, the training scripts themselves).\\
    \textit{Notable Instances:} While adding a new demo in a semantic segmentation project~\cite{train_infra_eg}, 12 files pertaining to the ML pipeline were changed. This was accompanied by a new training driver script\footnote{\url{https://github.com/TSchattschneider/PointCNN/commit/1827a79b2ede15007a06d327d95f10bc0753420}} for the new demo data. Clearly, the model building pipeline had to be updated at multiple places to accommodate this new data.\\

    \item \textbf{Pipeline Performance:}\\
    \textit{Definition:} Any change pertaining to the run-time efficiency of the ML pipeline. \\
    \textit{Explanation:} ML operations are computationally expensive and time-consuming. Iteratively retraining ML pipelines to find optimal values for model hyper-parameters and data cleaning configurations exacerbates performance needs. Hence, this sub-category of code changes involves any changes improving the run-time efficiency of the ML pipeline. \\
    \textit{Notable Instances:} Re-writing specific ML operations related to Principle Components Analysis (PCA) in Tensorflow    in a project~\cite{perf_eg} enabled higher computation efficiency. In particular, CPU utilization dropped from 5,600\% to 240\%.
\end{itemize}

\subsubsection{Meta Program}
As identified by Hindle et al., Meta Program\footnote{Note that Hindle et al.'s meta program change category is unrelated to the field of Metaprogramming (\url{https://en.wikipedia.org/wiki/Metaprogramming}).} changes update the metadata of the program (i.e., data required by the project, but not the source code). For instance, makefiles, and readme (\textit{external-documentation}) files.
We identified three new change sub-categories.

\begin{itemize}

    \item \textbf{Sharing.}\\
    \textit{Definition:} Changes in the way the source code of ML projects are presented or deployed to enable better collaboration between different roles involved in an ML project.\\
    \textit{Explanation:} In the modern collaborative development era, projects are shared with other developers and end users~\cite{zhou2020has}. In the case of ML pipelines, such changes involve converting python scripts into Jupyter notebooks better suited for understanding and working with complex ML operations~\cite{bloice2016tutorial}; or sharing the dependency environment via docker containers, enabling others to quickly deploy and run experiments on their infrastructure.\\
    \textit{Notable Instances:} A docker file was created for the project Neural-style~\cite{sharing_eg}. Another project changed the demo jupyter notebooks files~\cite{sharing_eg2} to disseminate the developed ML project and its parameters.\\

    \item \textbf{Validation Infrastructure.} \\
    \textit{Definition:} Changes made to the ML model validation component of an ML pipeline~\cite{8804457}.\\
    \textit{Explanation:} Validation changes involve changes to any modules or components responsible for driving the evaluation of a trained ML model's (accuracy) performance, possibly comparing to the performance of prior trained models or earlier iterations of the trained model. This kind of change is similar to \textit{Training Infrastructure} change, but focuses on the validation infrastructure instead of the training infrastructure. \\
    \textit{Notable Instances:} The file \texttt{evaluate3.py} was added in an image processing project~\cite{ch_ev_eg} to evaluate the model's performance by comparing the predicted labels (annotations on images) against the true labels. \\
    
    \item \textbf{Internal documentation}\\
    \textit{Definition:} Changes that explain the internal workings of the ML code to developers.\\
    \textit{Explanation:} Internal documentation changes clarify the fine-grained implementation of the code, with developers and data scientists as the intended audience. Such changes not only add code comments but could also add log statements to the code, for example, a succession of \texttt{print} statements, to better comprehend the workings of ML pipeline operations. Internal documentation changes differ from \textit{external documentation}, since the latter explicitly document a project for end-users, typically using README files or API documentation. \\
    We introduce \textit{internal documentation} as an augmentation to Hindle et al.'s~\cite{largecommits} \textit{Documentation} sub-category, as they did not provide any distinction between internal and external documentation.\\
    \textit{Notable Instances:} In a FasterRCNN project~\cite{comp_eg}, ambiguous internal documentation about an internal flag variable using \texttt{DEFINE\_bool} was rectified. In another project~\cite{comp_eg_2}, the grammar of the comments that explain the internal working of the code was fixed.
    
\end{itemize}

\subsubsection{Module management}

As identified by prior research, module management changes the way files are named and organized into source code modules. In addition to Hindle et al.'s sub-categories (i.e., \textit{add}, \textit{rename} and \textit{delete} module), we identified a new change sub-category, \textit{Adding auto-generated code}.
\begin{itemize}
    \item \textbf{Adding auto-generated code}\\
    \textit{Definition:} Adding new files to the project that are generated automatically by external tools, alternative IDEs, or varying environment configurations.\\
    \textit{Notable Instances:} A commit involving 8,491 added lines of code and 3,813 deleted lines of code across three C files~\cite{add_cython_nms_eg} corresponded to a re-generated C implementation of the Non-Maximum Suppression (NMS) algorithm, typically used for selecting the best bounding boxes of objects in an image. Since pure Python implementations of this algorithm are not scalable, data scientists tend to use the Cython dialect of Python, which allows generating efficient C code. 
\end{itemize}

\subsubsection{Data Category [NEW]}
Any change to the infrastructure that handles ingestion/egestion of domain-specific data (e.g., for training or testing) required by an ML pipeline, or to the metadata of said data (e.g., directory paths). Note that this category does not involve committing actual data files, since Git repositories are not the right place to store large-scale data. 
\begin{itemize}
    \item \textbf{Input Data.}\\
    \textit{Definition:} Code changes to the logic responsible for loading data or ingesting external data into an ML pipeline. \\
    \textit{Explanation:} 
    ML pipelines need to deal with a variety of data storage platforms (e.g., CSV files, SQL database, Kafka, data lakes) to obtain domain-specific input data. Hence, this sub-category of changes relates to the logic of dealing with such data platforms and the data schemas of ingested data.
    
    \textit{Notable Instances:} File \texttt{extract\_tfrecords\_main.py} in project, \texttt{Youtube-8m}~\cite{inp_eg}, added functionality to load external video frames data and feed it in the right data format to the ML pipeline.\\

    \item \textbf{Output Data.}\\
    \textit{Definition:} Changing the way the output data of the ML program is stored. \\
    \textit{Explanation:} Output data changes pertain to the way the results/output of the ML pipeline's are saved to the file system. Such changes may be needed to improve the integration of a model or its prediction results into an end-user application (e.g., UI applications or dashboards), or in other pipelines.\\
    \textit{Notable Instances:} The faster-rcnn demo program was changed to save its output to an image file~\cite{out_eg}.\\
    
    \item \textbf{Project Metadata.} \\
    \textit{Definition:} Changing the metadata 
    of all data files an ML project manages.\\
    \textit{Explanation:} ML pipelines contain a variety of metadata about the input and output data that they ingest/egest, such as paths of base directories or specific data files, license information of said data, etc.
    Hence, project metadata changes include adding, updating or deleting such metadata. This does not include changes to the actual data (\textit{pre-processing}) or the infrastructure used to ingest/egest such data (\textit{Input Data}/\textit{Output Data}), only to the project metadata. \\
    \textit{Notable Instances:} 
    Project directories for loading various model artifacts like model graphs and pretrained models were updated in a facenet implementation~\cite{prog_eg}.
\end{itemize}
 
\subsubsection{Source Management}
    Hindle et al. described \textit{Source management} as changes performed due to the way a version control system is being used by a project. Along with the five sub-categories identified by Hindle et al.~\cite{largecommits}, we identified one new sub-category.
    \begin{itemize}
    \item \textbf{Changing file permissions.}\\
    \textit{Definition:} Changes adding, updating, or removing file permissions (like executability of a script).\\
    \textit{Explanation:} Traditional programs and ML operations are often run on shared high-performance computers (typically Unix-based servers). Managing file permissions is essential for assigning ownership of files while dealing with multiple users, thereby enforcing security.
    
    \textit{Notable Instances:} The file start.sh was given 775 permissions~\cite{ch_fi_eg} since it's previous 664 permission did not allow the script to be executed by the owner of the file or its Unix user group.
    \end{itemize}

\subsubsection{Dependency Management [NEW]}
    While we identified this new category related to handling third-party dependencies (e.g., libraries or packages of a Linux distribution) on code changes of the studied ML pipelines, the management of such dependencies is common across both ML and non-ML projects~\cite{decan2019empirical,pashchenko2020qualitative,mukherjee2021fixing}.

\begin{itemize}

    \item \textbf{Add Dependency.} \\
    \textit{Definition:} Adopting a new third-party dependency in the source code.\\ 
    \textit{Explanation:} Adoption of a new third-party dependency typically requires adding the name and version of the dependency to a configuration file, as well as adding import statements to various files in the source code, in order to declare the dependency to compilers or interpreters.\\
    \textit{Notable Instances:} Addition of new import statements like \\
    ``\texttt{from tensorflow.python.lib.io import file\_io}''~\cite{add_pkg_eg}.\\

    \item \textbf{Remove Dependency.} \\
    \textit{Definition:} Stopping the adoption of a third-party dependency.\\
    \textit{Explanation:} Removing an unused import statement from a source code file, or even removing the actual third-party dependency from the list of dependencies of a file. \\
    \textit{Notable Instances:} Removal of unused import statements~\cite{add_pkg_eg}.\\

    \item \textbf{Update Dependency.}\\
    \textit{Definition:}After the adoption of a dependency, changes might be needed to the metadata of the dependency.\\
    \textit{Explanation:} This change category involves updating the metadata of a dependency, for example, to keep the dependency compatible with the code base, or vice versa. This typically includes updating the dependency version. \\ 
    \textit{Notable Instances:} Change of the \texttt{cloudml-gpu} runtime version from ``1.0'' to ``1.8''~\cite{update_pkg_eg}.
\end{itemize}
\begin{figure}[t]
    \hspace{-0.7cm}
    \includegraphics[width=1.1\columnwidth,keepaspectratio]{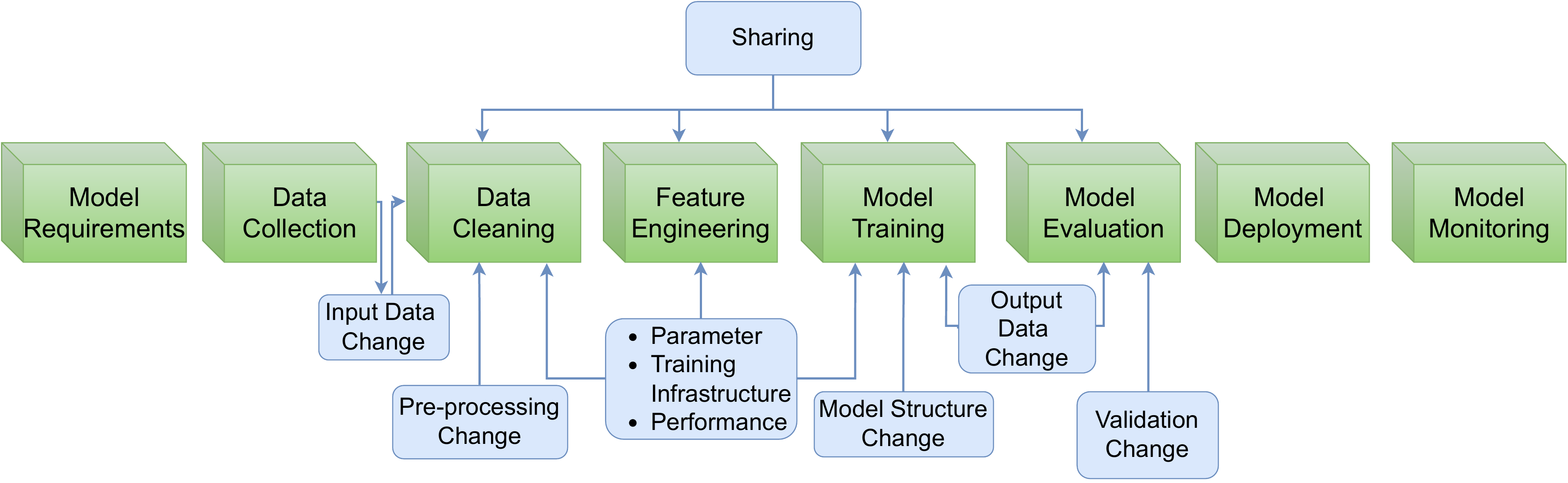}\caption{ML-specific sub-categories mapped to Amershi's ML pipeline architecture~\cite{8804457}.}
    \label{fig:amershi_bhatia_map}
\end{figure}

\subsubsection{Mapping the updated change taxonomy to Amershi's ML pipeline architecture}

In Figure~\ref{fig:amershi_bhatia_map}, we provide an association between the nine new ML-specific categories of code changes identified in this research to the ML pipeline architecture of Amershi et al.~\cite{8804457}. We notice that most (6) of the identified change categories apply to the \textit{data cleaning} and \textit{model training} phases, followed by the \textit{feature engineering} (4) and \textit{model evaluation} (3) phases.\\
On the other hand, none of the change types map to the initial phases of \textit{model requirements}, and \textit{data collection}, since those involve tasks performed by management and data engineers, respectively.
Similarly, the end phases, namely, \textit{model deployment} and \textit{model monitoring}, are geared towards third-party applications where the trained model is integrated and deployed by MLOps engineers into (amongst others) dashboards, UI applications, and back-end servers for prediction.

\begin{Summary}{Summary of RQ2}{}
Hindle et al.'s taxonomy of software code changes~\cite{largecommits} had to be extended with two high-level change categories (ML-specific \textit{Data}, and generic \textit{Dependency management}). We also extended the taxonomy by identifying 16 new sub-categories of changes, nine of which (i.e., \textit{input data, parameter tuning, pre-processing, training infrastructure, model structure, pipeline performance, sharing, validation infrastructure, and output data}) are ML-specific.
\end{Summary}
\begin{figure}[t]
    \centering
    \includegraphics[width =1.2\columnwidth,keepaspectratio]{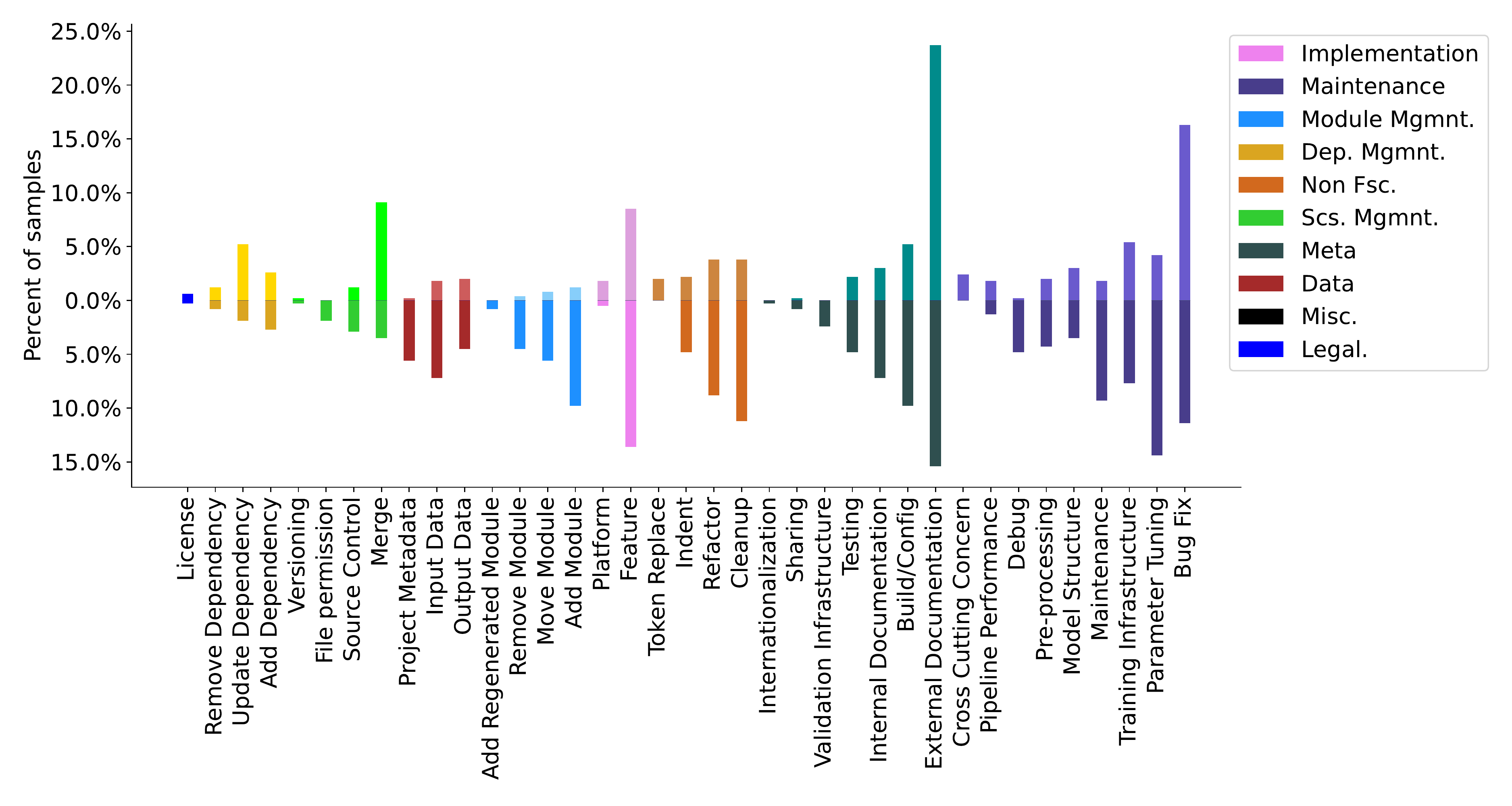}
    \caption{Percentage of change sub-categories present in the 378 samples of Downstream commits and 539 samples of Upstream commits. Values above $y=0.0\%$ (plotted using a lighter color palette) indicate the percentage of upstream changes containing a specific change sub-category, whereas values below $y=0.0\%$ (darker color palette) indicate the same for downstream changes.}
    \label{fig:up_vs_down}
\end{figure}\label{subsec.RQ2}
\subsection{\rqthree}

\noindent\textbf{Motivation.} 
In \textbf{RQ1}, we observed that 41.6\% of \texttt{Forks\_with\_changes} submit upstream contributions back to the ML research repositories. Since this means that contributions by more than half of the forks were never sent upstream, it is interesting to understand the nature of such contributions, i.e., what did the ML community need in addition to the original development in the parent repository (merged PRs), and what did the authors of the ML repository miss out on (i.e., code changes not contributed back)? In particular, the upstream changes studied in this paper help to identify the missing aspects of the original ML parent repository contributed back by the OSS community. Conversely, understanding the downstream changes studied in this paper helps us determine to what extent essential features or contributions have been missed.

\noindent\textbf{Approach.} 
This RQ uses the sample of 378 downstream changes and 539 upstream changes labeled with high inter-rater agreements in \textbf{RQ2}, but this time to analyze the prevalence of each change sub-category of the taxonomy in Figure~\ref{fig:Finaltaxonomy}. In particular, we compute the percentage of upstream and downstream changes for each (sub-)category, then compare our findings between downstream and upstream changes. These results are summarized in Figure~\ref{fig:up_vs_down}.

\noindent\textbf{Results.} 
\textbf{For both upstream and downstream commits, heavy changes occur in \textit{Documentation} (Meta program), \textit{Bug fixes} and \textit{Model training} (Maintenance); and adding new \textit{features} (Implementation).} Figure~\ref{fig:up_vs_down} shows substantial peaks in the \textit{Maintenance} and \textit{Meta program} categories. We attribute such results to the nature of the data science life cycle, where ML pipelines require substantial \textit{maintenance} activities during experimentation with and tweaking of data and models. ML tasks include multiple iterations of updating data \textit{pre-processing,} tuning \textit{parameters,} updating \textit{model building} code, and improving \textit{pipeline performance}.

\textbf{Maintenance changes are much more prevalent in downstream commits than in upstream commits.} This is visible through the higher percentages of \textit{non functional} and \textit{data} changes in downstream commits.  We attribute this imbalance to downstream users adapting ML research for their domain-specific tasks, rather than improving the upstream repository for generic usage. In order to improve personal understanding of the ML code, downstream users added changes from the new change sub-category, \textit{internal documentation}, such as added $print$ statements. In contrast, we found no cases of \textit{internal documentation} in upstream commits.

We also observed nine cases of our new sub-category \textit{change evaluation} for downstream changes, while none for upstream commits. Intuitively, downstream developers needed scripts to train and test models (potentially after making some other changes) against their domain-specific data. Conversely, there were 45 cases of our new change sub-category, \textit{Dependency Management}, in upstream commits. Such changes \textit{add, update} or \textit{remove} ML library dependencies. Finally, accepted PRs were merged into either a project's main branch or alternative branches, which led to more instances of \textit{merging} changes (Source Management) in upstream commits than in downstream commits.

\paragraph{\textbf{Noticeable instances of the most popular pre-AI change categories.}\\} 

In the results of RQ3, we notice that 3 of the top 4 most occurring change types across upstream and downstream commits, i.e., Bug Fix, Documentation, and Feature, belong to Hindle's pre-AI taxonomy. Hence, here we provide some examples of those change sub-categories in the context of ML-based pipeline projects. 
\begin{enumerate}
    \item External Documentation:  In the project \textit{TensorBox}, the README.md file was updated to present information about how to manually download external dependencies (e.g., CUDA version) and how to set up and configure the corresponding runtime environment~\cite{doc_eg}. This was a downstream change. 
    \item Bug Fix: In the project \textit{tf-faster-rcnn}, a bug in the \texttt{test\_rpn} function was fixed in a downstream change as indicated by its commit message~\cite{bug_fix_eg}. In another project, \textit{Kitti-Seg}, the order of height and width parameters was incorrectly swapped, as indicated by the commit message for the upstream change~\cite{bug_fix_eg2}.
    \item Feature: In a downstream commit for project \textit{PointCNN}~\cite{fea_eg}, new features were added allowing to set GPU flags and the CUDA path, and to create project directories for saving the model, if not yet existing.
\end{enumerate}

\begin{Summary}{Summary of RQ3}{}
Both upstream and downstream contributions to ML research repository add \textit{documentation}, \textit{fix bugs} and \textit{add features}. Downstream developers change \textit{input/output data}, perform \textit{parameter tuning}, add new functional \textit{features}, and perform other non-functional changes like \textit{indentation}, \textit{refactoring}, or \textit{cleaning up} the source code. Such changes are domain oriented. On the other hand, upstream changes benefit the parent repository by \textit{updating dependencies} or \textit{fixing bugs} for the parent repository.
\end{Summary}
\label{subsec.RQ3}
\section{Implications}\label{sec.impl}
In this section, we discuss the implications of our findings for software researchers, ML practitioners, the OSS ML community, toolsmith engineers, and ML educators.

\subsection{Implications for Researchers}
We extended Hindle et al.'s taxonomy of code changes~\cite{largecommits} with two new categories and 16 new sub-categories of changes. \textbf{Researchers can use our extended taxonomy to obtain a holistic picture of software changes in ML pipelines}. In particular, nine (\textit{input data, parameter tuning, pre-processing, training infrastructure, model structure, pipeline performance, sharing, validation infrastructure, and output data}) of the ML-related categories of code changes indicate a need for revising and adapting existing best practices towards the needs of software engineering for ML systems. This is only exacerbated by the prevalence of the \textit{internal} and \textit{external documentation} change sub-categories, indicating difficulties for developers to comprehend complex ML code and keep track of hefty ML pipelines.

\textbf{Our work also updates existing code-change dimensions towards modern SE paradigms in SE4AI}. Even though some code change categories identified by Hindle et al. still apply in the context of ML pipelines, we were able to better understand their applicability within our context of ML pipelines. For instance, studying software bugs has been an important focus of the software engineering community for decades. With the advent of ML, recently many studies ~\cite{chen2022maat,tizpaz2020detecting,cheng2018manifesting,dwarakanath2018identifying} started focusing on bugs in the machine learning domain. However, thus far the scope of these types of ML studies is 
limited to machine learning frameworks, while bugs in the different phases and components of actual ML pipelines or even end-user ML applications are not yet explored in depth. For example, initial studies found that 
the data wrangling phase 
introduces a variety of pipeline-level~\cite{8804457} challenges, including pipeline-level bugs.

As another example, we split the ``documentation'' category of Hindle's change taxonomy into ``internal'' and ``external'' documentation, since our analysis of ML pipeline code changes made it especially apparent that both cater to a different audience in the modern SE paradigm of collaborative development. In particular, the (external) API-level or application-level documentation is aimed toward black-box (re)use of a given project, while the fined-grained (internal) documentation instead explains the inner working and state of processing of the code to people interested in changing, or at least better understanding it. Such a distinction may not have been that obvious 13 years ago~\cite{largecommits}.

\subsection{Implications for Toolsmiths}
\textbf{An updated taxonomy of code changes can help toolsmiths in adapting and innovating software engineering tools.} As mentioned in Section~\ref{sec.related}, code change data is used for a variety of purposes such as extraction of missing traceability links~\cite{wu2011relink}, auto-generation of commit messages~\cite{6975661}, and analysis of quality impact~\cite{farago2014impact}. At the same time, current development environments and tools used by developers need to be modernized as well. 

Our observed instances of code changes spanning across the \textit{Maintenance, Dependency Management, Source management}, and \textit{Data} domains imply a need for toolsmiths to 
\textit{better support ML engineering teams in handling requirements, managing data dependencies, configurations, training ML models.} For example, given that many developers add comments to the code to better understand the ML logic, code bases might get polluted. The boom of Jupyter Notebooks~\cite{granger2021jupyter} for data science only provides a workaround to this problem~\cite{wang2020better}, which might not scale to real-life ML practices of large systems. Hence, perhaps less invasive annotation or other functionality is required in future IDEs.

As another example, code changes play an important role in tracking bugs~\cite{zhao2017towards,kim2011empirical,shivaji2012reducing}. An updated taxonomy of code changes may enable a more effective automated classification of bug reports or code changes submitted for code review. For example, most automated change classification techniques focus on a limited number of possible change types. Our results could also help fault localization and defect prediction researchers improve their models. 
\\


\subsection{Implications for Educators}
Software educators may wish to update their curricula to revise future training of (ML) software engineers. Moreover, novice software developers need training on practices related to \textit{Dependency management}, to be better equipped to use and support ML frameworks. Overall, ML practitioners need to be aware of the change taxonomy to anticipate future changes that occur in ML software.

Particularly, Figure~\ref{fig:amershi_bhatia_map} provides a map for educators of the different ML-related change (sub-)types to expect while providing a travel guide for education and training teams. Since ML-based organizations tend to have a distributed team with overlapping roles ranging from data developers, data scientists, statisticians, DevOps engineers, to software developers, software teams may wish to leverage such a map for a clearer understanding of the roles and responsibilities w.r.t. the nature of development changes performed by a specific role.

\subsection{Implications for OSS community}
\textbf{Organizations and/or individuals wishing to open-source their ML repositories should have realistic assumptions.} While our findings show that organizations and researchers do not necessarily ``dump'' their ML research implementations on GitHub, but receive and merge open-source contributions, this is not guaranteed. For one, only 9\% of forks are \texttt{Forks\_with\_changes}, of which 41.6\% send contributions upstream via a pull request, about half of which (52.1\%) are accepted into the parent repository (see \textbf{RQ1}).

Two lessons can be learned from this. On the one hand, the ML research repositories are missing out on almost 60\% of forks having contributions that are never sent back upstream. Even the 41.6\% of forks that do contribute might not contribute back all contributions they have made. While it is OK for changes like \textit{parameter tuning} not to be contributed back, the ``lost'' contributions of forks also include 16\% of new bug fixes, 13\% of new features for the ML pipeline, etc. Future work should look into why those were never sent back.

On the other hand, of those contributions that were propagated back, only half were merged. Future work should consider the reasons for rejection of this work, i.e., to what extent was rejection based on the quality of the contribution versus the contribution being too tied to the contributor's own use case, or even versus the responsiveness of the ML repository owners. Whichever the outcome, and similar to traditional open-source development, receiving many high-volume contributions requires effort~\cite{rahman2014insight}. Researchers can gain leverage from our updated taxonomy to pay special attention to ML pipeline components that are updated while maintaining ML code.
\section{Threats to Validity}\label{sec.threats}

\textbf{Threats to Internal Validity.} Qualitative studies can be subject to researcher bias. To minimize this, we used multiple participants (i.e., two teams with four people in Team-A, and three people in Team-B) for the manual coding of changes. Both teams had in-depth knowledge of software development, as well as SE4ML. Furthermore, the teams pair-wise labeled each sample and achieved high inter-rater agreements of Krippendoff's $\alpha = 98\%$ for Team-A and $\alpha = 92\%$ for Team-B.\\

\noindent\textbf{Threats to Construct Validity.} In Section~\ref{sec.data}, we mine ML pipeline repositories implementing algorithms published in ArXiv papers. For this, we check whether a repository cites an ArXiv research paper in its README file. Analysis of a sample of 94  such repositories in Section~\ref{fig.datacollection} showed that all repositories citing an ML ArXiv publication are influenced by the research and can be termed as a ``ML research repository'', irrespective of whether the repository is created by the authors of the ML research publication (30\%) or by external members of the community (70\%).\\

\noindent\textbf{Threats to External Validity.} For answering our \textbf{RQ2}, we analyzed both the types of changes made by PRs merged into the upstream parent repository and by downstream changes performed within the forks but never sent upstream. However, we do not study changes rejected by the upstream repository. While future work should analyze such cases, we feel confident about the completeness of our taxonomy, as we reached saturation in obtaining new labels within the initial 78 samples of downstream commits. No new categories were found in the later part of 300 downstream or any of the 539 upstream commits.

Moreover, we focused on the repositories implementing image processing or machine learning in ModelDepot, since they were the most popular on Modeldepot and cover a wide range of popular ML application domains. Future work should focus on other domains like NLP and Audio Processing. 

Finally, we sampled our data for qualitative analysis only from the 23 repositories that were at the top five percentile of \textit{Forks\_with\_changes}. We put such a filter to select repositories with the maximum amount of activity in terms of downstream commits and pull requests which may thereby manifest as upstream commits.
While we need ``popular or active trends'' to study rich and meaningful data that has low noise, nonetheless, this poses a threat to external validity as is also indicated by prior research~\cite{Kalliamvakou14,santos2015investigating,bird2009fair}.\\
\noindent\textbf{Threats to Reliability Validity.} These threats take into account the replication of our study. After our data collection process from ModelDepot finished and the analysis was well underway, the website was shut down. However, ModelDepot only pointed to the repositories hosted on GitHub. To mitigate this threat, we provide\footnote{\url{https://github.com/SAILResearch/suppmaterial-22-aaditya-ml_change_taxonomy}}
 our lists of 1,346 repositories, along with the ArXiv papers cited by these repositories.
We also provide a snapshot of the mined forking data at the time of our analysis for our quantitative investigation of \textbf{RQ1}. The replication data for qualitative analysis of \textbf{RQ2, RQ3} consists of the labeled sample of upstream and downstream changes to ensure the reproducibility of our study.
\section{Conclusion}\label{sec.conclusion}
Open-source community developers use and refine OSS repositories. Although prior studies have investigated the nature of open source contributions in non-ML software, one can imagine the nature of such community changes, as well as the way in which developers collaborate, to be different for Machine Learning projects. Hence, this paper studies the forking dynamics and the types of changes performed in 67,369 forks of 1,346 ML pipeline projects related to research publications.

We found that, while most forks (91\%) do not modify an ML research repository after forking it, 41.6\% of the forks with modifications contribute valuable changes to the parent ML research repository, with a 52.1\% acceptance rate. We performed an extensive qualitative study that identified the types of changes in ML software. We identified one new top-level change category, \textit{Data}, in the context of ML, and one more generic category (\textit{Dependency management}). Along with this, we extend the taxonomy of changes by adding 15 new sub-categories, including nine ML-specific ones (\textit{input data, output data, program data, sharing, change evaluation, parameter tuning, pipeline performance, pre-processing, model training}) and seven generic ones (i.e. \textit{adding dependency, removing dependency, updating dependency, file permissions, internal-documentation, adding auto-generated code} and \textit{project metadata}).

Our results aim to help software practitioners in having a better understanding of ML changes that can be leveraged while training new developers, and to support building and maintaining ML software. Furthermore, future work should look deeper into the reasons why potentially valid \textit{Documentation}, \textit{Feature} and \textit{Bug fix} changes were not contributed back upstream.

\section{Data Availability}
All the scripts along with the mined data are provided in the replication package\footnote{\url{https://github.com/SAILResearch/suppmaterial-22-aaditya-ml_change_taxonomy}}

\section*{Acknowledgement}
We thank Greg Wilson for providing insightful ideas and comments for this work. We also thank Boyuan Chen, Minke Xiu, Javier Rosales and Wanqing Li for their contributions to the analysis and feedback on this work. 
\bibliographystyle{IEEEtran}
\bibliography{references}

\end{document}